\documentclass[english,twocolumn,showpacs,preprintnumbers,amsmath,amssymb]{revtex4}
\usepackage[T1]{fontenc}
\usepackage[latin9]{inputenc}
\usepackage{amsmath}
\usepackage{graphicx}
\usepackage{amssymb}

\makeatletter

\providecommand{\tabularnewline}{\\}

\@ifundefined{textcolor}{}
{%
 \definecolor{BLACK}{gray}{0}
 \definecolor{WHITE}{gray}{1}
 \definecolor{RED}{rgb}{1,0,0}
 \definecolor{GREEN}{rgb}{0,1,0}
 \definecolor{BLUE}{rgb}{0,0,1}
 \definecolor{CYAN}{cmyk}{1,0,0,0}
 \definecolor{MAGENTA}{cmyk}{0,1,0,0}
 \definecolor{YELLOW}{cmyk}{0,0,1,0}
 }

\makeatother

\usepackage{babel}

\begin{document}

\title{Quadrupole collectivity in random two-body ensembles}

\author{Volha Abramkina and Alexander Volya}

\affiliation{Department of Physics, Florida State University, Tallahassee, FL
32306-4350, USA}

\date{\today}
\begin{abstract}
We conduct a systematic investigation of the nuclear collective dynamics
that emerges in systems with random two-body interactions. We explore
the development of the mean field and study its geometry. We investigate
multipole collectivities in the many-body spectra and their dependence
on the underlying two-body interaction Hamiltonian. The quadrupole-quadrupole
interaction component appears to be dynamically dominating in two-body
random ensembles. This quadrupole coherence leads to rotational spectral
features and thus suggests the formation of the deformed mean-field
of a specific geometry. 
\end{abstract}



\maketitle

\section{Introduction}

Emergent phenomena is one of the most profound topics in modern science
addressing the ways that collectivities and complex patterns appear
from multiplicity of components and simple interactions. Ensembles
of random Hamiltonians allow one to explore the emergent phenomena
in a statistical way, and thus to establish generic relations and
rules. To study the many-body physics of interest we adopt a shell
model approach with a two-body interaction Hamiltonian. The sets of
the two-body interaction strengths are selected at random resulting
in the two-body random ensemble (TBRE). Symmetries, such as rotational,
isospin, and parity, entangled with complex many-body dynamics result
in surprising regularities discovered recently in the low-lying spectrum.
Patterns exhibited by the random ensembles are remarkably similar
to those observed in real nuclei. The high probability for the ground
state spin to be zero is the most astounding feature of the TBRE discovered
in Ref. \cite{Johnson:1998}. Signs of almost every collective feature
seen in nuclei, namely, pairing superconductivity, deformation, and
vibration, have been observed in random ensembles \cite{Zelevinsky:2004,Johnson:2007,Papenbrock:2007,ZhaoPR:2004,Bijker:2000}.
While the systematics of the ground state quantum numbers is almost
not sensitive to the short-range pairing matrix elements, the probability
to find a coherent paired structure in the wave-functions of low-lying
states is enhanced \cite{Zelevinsky:2004}. The presence of rotational
features in the spectra is another unexpected result seen in the TBRE \cite{Zelevinsky:2004,Horoi:2010}. 

The goal of this work is to study the emergence of collective mean-field
dynamics in ensembles with random interactions. The discussion is
organized as follows: In Sec. \ref{sec:Collectivity} we briefly define
the TBRE, introduce signatures of collective motion, and discuss ways
to detect them. In Sec. \ref{sec:single_level} we present our study
of collectivities in single$j$ level models. More complex models
are explored in Secs. \ref{sec:two_levels} and \ref{sec:Realistic-Model-Space}.
We summarize our results in Sec \ref{sec:Summary} with a discussion
of the quadrupole-quadrupole Hamiltonian which appears to be responsible
for most of the observed phenomena.

\section{Collective observables and models\label{sec:Collectivity}}

In the spirit of the traditional shell model approach, we define a
model configuration as $(j_{1},\, j_{2}\dots)^{N},$ where $N$ nucleons
occupy a set of single particle levels labeled by their angular momentum
$j$. In this work we assume that the single particle energies are
degenerate. We examined other models for which this was not the case
and the results are similar. The Hamiltonians in the TBRE are defined
with a set of two-body matrix elements which are selected at random.
The distribution of the matrix elements is Gaussian so that, within
a given symmetry class, the ensemble of Hamiltonian matrices for two
particles coincides with Gaussian Orthogonal Ensemble. The presence
of rotational symmetry and, where relevant, of parity and isospin
symmetries is assumed. The typical number of random realizations was
between $10^{5}$ and $10^{7}$ for all ensembles presented in this
work.

In the TBRE the number of realizations where the ground state spin
$J_{gs}=0$ is disproportionally large. Aiming at collective phenomena
we select realizations with $J_{gs}=0$. With the exception of the
ground state, labeled as $0_{gs}$, we denote the low-lying states
by the value of their spin with an identifying subscript. The subscript
is given in bold font if it refers to the absolute position of a given
state in the spectrum. Throughout the paper we give probabilities
of finding realizations with certain features, these probabilities
are always quoted in percent relative to the size of the ensemble;
however, all probability distribution plots are normalized to unit
area. 

In order to identify and to analyze manifestations of collective phenomena
in the spectra we use a set of observables. The goal is to choose
a finite number of spectral observables that are likely to convey
the most information about possible collective structures in a scale-independent
way and with minimal model dependence. These quantities and the logic
behind their selection are discussed in what follows.

The geometry of the nuclear mean field is described by the multipole
density operators ${\cal M}_{\lambda\mu}$ with multipolarity $\lambda$
and magnetic component $\mu$. The structure of the multipole operators
depends on the valence space, for each model it is addressed separately.
The reduced transition rate from an initial state $|JM\rangle$ to
a final state $|J'M'\rangle$ \begin{equation}
B(E\lambda,J\rightarrow J')=\frac{1}{2J+1}\sum_{\mu,M,M'}|\langle J'M'|\mathcal{M}_{\lambda\mu}|JM\rangle|^{2}\label{eq:BE}\end{equation}
is one of the observables. Here $|JM\rangle$ denotes a many-body
state with angular momentum $J$ and magnetic projection $M.$ The
total transition strength from a state $J$ is given by the sum rule
\begin{equation}
S_{\lambda}(J)=\sum_{J'}B(E\lambda,J\rightarrow J')=\langle JM|\sum_{\mu}\mathcal{M}_{\lambda\mu}^{\dagger}\mathcal{M}_{\lambda\mu}|JM\rangle,\label{eq:QB}\end{equation}
which provides a convenient normalization to assess the \emph{fractional
collectivity} of the transition \begin{equation}
b(E\lambda,J\rightarrow J')=\frac{B(E\lambda,J\rightarrow J')}{S_{\lambda}(J)}.\label{eq:b}\end{equation}

The shape of a state is described by its multipole moments specified
by the expectation value \begin{equation}
{\cal Q}_{\lambda}(J)=\langle JJ|\mathcal{M}_{\lambda0}|JJ\rangle.\label{eq:Qlab}\end{equation}
For a non-spherical system this moment describes the shape of a deformed
nucleus measured in the lab frame. The intrinsic shape is characterized
by the body-fixed (intrinsic) multipole moments $Q_{\lambda}.$ A
rotational spectrum (band) emerges for every fixed intrinsic shape.
In a rigid rotor these intrinsic moments are the same for all states
in the band and they determine the lab-frame observables in Eqs. (\ref{eq:BE})
and (\ref{eq:Qlab}). For the ground state band of interest, the intrinsic
moments determine the total transition strength $S_{\lambda}(0_{gs})=Q_{\lambda}^{2}$. 

In the axially symmetric case the quantum number $K$, a projection
of the angular momentum onto the body-fixed symmetry axis, is conserved.
Then for each rotational $K$-band the relations between the observables
in the lab frame and in the intrinsic frame are expressed via Clebsch-Gordan
coefficients ${\cal Q}_{\lambda}(J)=Q_{\lambda}C_{\lambda0,JJ}^{JJ}C_{\lambda0,JK}^{JK}$
and $B(E\lambda,J\rightarrow J')=Q_{\lambda}^{2}\left|C_{\lambda0,JK}^{J'K}\right|^{2}.$
This limit of an axially symmetric rotor provides a convenient normalization
to examine the multipole moments. In this work instead of ${\cal Q}_{\lambda}(J)$
we quote a normalized intrinsic moment \begin{equation}
q_{\lambda}(J)=\frac{Q_{\lambda}(J)}{\sqrt{S_{\lambda}(0_{gs})}},\,{\rm where}\quad Q_{\lambda}(J)=\frac{{\cal Q}_{\lambda}(J)}{C_{\lambda0,JJ}^{JJ}C_{\lambda0,J0}^{J0}},\label{eq:q}\end{equation}
which is computed as if the state is a member of the $K=0$ rotational
ground state band. 

In this paper we only briefly touch the subject of collectivities
other than quadrupole, see Sec. \ref{sub:Higher-multipole-collectivity};
thus for convenience the subscript $\lambda$ is omitted for $\lambda=2$.
The relation between the lab-frame moment of the $2_{1}$ state and
its intrinsic moment is ${\cal Q}(2_{1})=-2/7\, Q(2_{1}).$ For the
axially symmetric rotor the quadrupole transition sum rule for the
$0_{gs}$ is saturated by a single transition $b(E2,0_{gs}\rightarrow2_{1})=1.$
The quadrupole moment is $q(2_{1})=1$ for prolate or $q(2_{1})=-1$
for oblate\emph{ }shapes. 

We normalize the total transition strength $S_{\lambda}(J)$ to its
maximum possible value for a given valence space. Taking the $\lambda=2$
case as an example, we define the quadrupole-quadrupole (QQ) Hamiltonian
as \begin{equation}
H_{{\rm QQ}}=-\sum_{\mu}\mathcal{M}_{2\mu}^{\dagger}\mathcal{M}_{2\mu}.\label{eq:QQHAM}\end{equation}
The eigenstate energy of the QQ Hamiltonian coincides with the total
transition strength \eqref{eq:QB} for that state: $E_{\text{QQ}}(J)=-S(J)$.
Thus, the absolute value of the ground state energy of the QQ Hamiltonian
$\left|E_{\text{QQ}}(0_{gs})\right|$ is the maximum possible value
of the total transition strength $S(J)$ for a given model space and
for a given structure of the quadrupole operator. We therefore define
a \emph{relative transition strength} as\begin{equation}
s(J)=\frac{S(J)}{\left|E_{{\rm \text{QQ}}}(0_{gs})\right|}.\label{eq:s}\end{equation}

To summarize, in our study we use the dimensionless variables defined
in Eqs. (\ref{eq:b}), (\ref{eq:q}), and (\ref{eq:s}). To shorten
notations we define $b\equiv b(E2,0_{gs}\rightarrow2_{1}),\, q\equiv q(2_{1}),$
and $s\equiv s(0_{gs}).$ For collective models of pairing, rotations,
and vibrations $b\approx1.$ We refer to a realization with $b>0.7$
as \emph{collective} and with $b<0.3$ as \textit{non-collective}.
The quadrupole moment $q$ allows one to separate different collective
modes: $q\approx\pm1$ for rotations and $q\approx0$ for vibrations
and for paired states. In what follows we allude to collective realizations
with $q>0.7$ as \emph{prolate} and those with $q<-0.7$ as \emph{oblate.}
For rotations the relative transition strength $s$ is proportional
to the square of the intrinsic moment, and thus it is associated with
the Hill-Wheeler deformation parameter $\beta^{2}$. Within Elliot's
SU(3) model \cite{Elliott:1958} the relative transition strength
$s$ can be thought to represent the expectation value of the Casimir
operator which identifies the irreducible representation. In cases
where $s\approx1$ the ground state band structure is close to that
of the QQ Hamiltonian. 

The collective structure is further analyzed using the following $4_{1}$
state. The types of collective modes can be classified by the ratio
of the excitation energies measured relative to the energy of the
$0_{gs}$ state \begin{equation}
R_{42}=\frac{E(4_{1})}{E(2_{1})}.\label{eq:RR}\end{equation}
 This ratio is close to $0$ for pairing, 2 for vibration, and 10/3
for rotation. The ratio of deexcitation rates\begin{equation}
B_{42}=\frac{B(E2,4_{1}\rightarrow2_{1})}{B(E2,2_{1}\rightarrow0_{gs})}\label{eq:DX}\end{equation}
is another measure. It is nearly 0 for pairing, 2 for vibrational
mode, and 10/7 for rotational motion. Typically, for models with the
QQ Hamiltonian $R_{42}$ and $B_{42}$ are close to the rotational
values, see summary in Tab. \ref{tab:QQ}. A comprehensive review
of different collective models, their analytic predictions, and comparisons
with rotational spectra observed in real nuclei can be found in the
textbooks \cite{Casten:2000,Iachello:1987}.

\section{The single $j$ level model\label{sec:single_level}}

\subsection{Quadrupole collectivity\label{sub:Quadrupole-collectivity}}

We begin our presentation with single $j$ level models. Starting
from the original paper \cite{Johnson:1998} the single $j$ level
with identical nucleons has been at the center of numerous investigations;
a good summary may be found in the following reviews \cite{Zelevinsky:2004,Papenbrock:2007,ZhaoPR:2004,Huu-Tai:2002}.
With many issues understood and with still unanswered questions, the
single $j$ model remains an important exploratory benchmark. The
model, while simple, has a number of particularly attractive features
which can be of both advantage and disadvantage \cite{Volya:2002PRC}:
the Hamiltonian is defined with a small number of parameters; apart
from an overall normalization constant, the multipole operators are
uniquely defined; a special role is played by the quasispin SU(2)
group; and the particle-hole symmetry is exact. 

\begin{figure}
\includegraphics[width=3.4in]{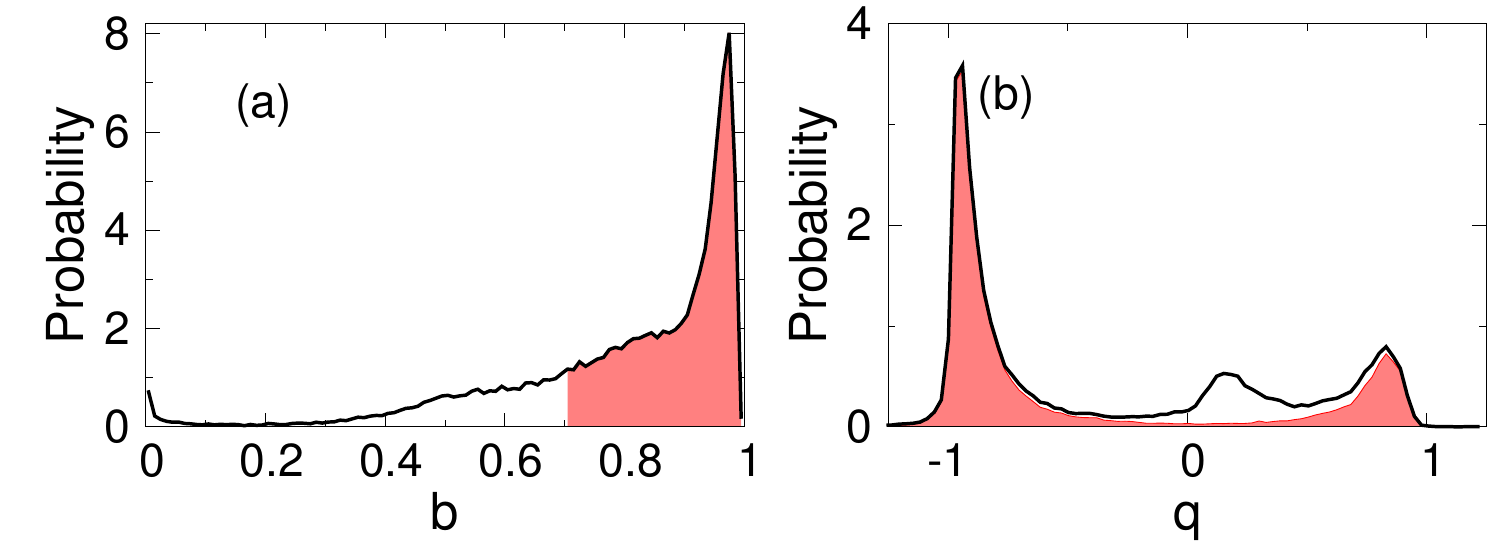}

\caption{(Color online) $(19/2)^{6}$. (a) The distribution of the fractional
collectivity $b$. (b) The distribution the intrinsic quadrupole moment
$q$. Only realizations with the $0_{gs},2_{\boldsymbol{1}}$ spin
sequence are included in both panels. There are 10.4\% of such realizations.
The 7.8\% of collective realizations $(b>0.7)$ are shaded. }
\label{fig:jl19N6}
\end{figure}

In Fig. \ref{fig:jl19N6} the system with 6 nucleons in a single $j=19/2$
level is examined, we refer to this system as $(19/2)^{6}$. Here
we select 10.4\% of random realizations where the $0_{gs}$ state
is followed by the $2_{\boldsymbol{1}}$ state. The distribution of
the fractional collectivity $b\equiv b(E2,0_{gs}\rightarrow2_{1})$
in Fig. \ref{fig:jl19N6}(a) points to highly collective nature of
the quadrupole transition $0_{gs}\rightarrow2_{\boldsymbol{1}}.$
Most realizations with $0_{gs}$ and $2_{\boldsymbol{1}}$ are collective
$(b>0.7)$, their fraction is 7.8\% of the total number of samples.
These realizations are shaded in red. This collectivity is not a statistical
coincidence. The system $(19/2)^{6}$ has 1242 spin-states, among
them there are 10 states with $J=0$ and 23 states with $J=2$. Thus,
statistically the chance for the $0_{gs},2_{\boldsymbol{1}}$ spin
sequence to occur among all other possible outcomes is only 0.015\%.
The large fractional collectivity for the transition between these
two states is even more unlikely, given that the transition strength
is shared among 23 $J=2$ states, the chances for $b(E2,0_{gs}\rightarrow2_{1})>0.7$
are of the order of 1 in $10^{7}$. 

There are two peaks in the distribution of the quadrupole moment in
Fig. \ref{fig:jl19N6}(b), they reflect prolate and oblate deformations.
For most of the collective realizations, which are shaded in Fig.
\ref{fig:jl19N6}, the magnitude of the quadrupole moment is consistent
with the value for the axially deformed rigid rotor $(|q|\thickapprox1)$.
The ground state is most likely to be oblate, but in about one out
of four collective cases a prolate mean field emerges. 

\begin{figure}
\includegraphics[width=3.4in]{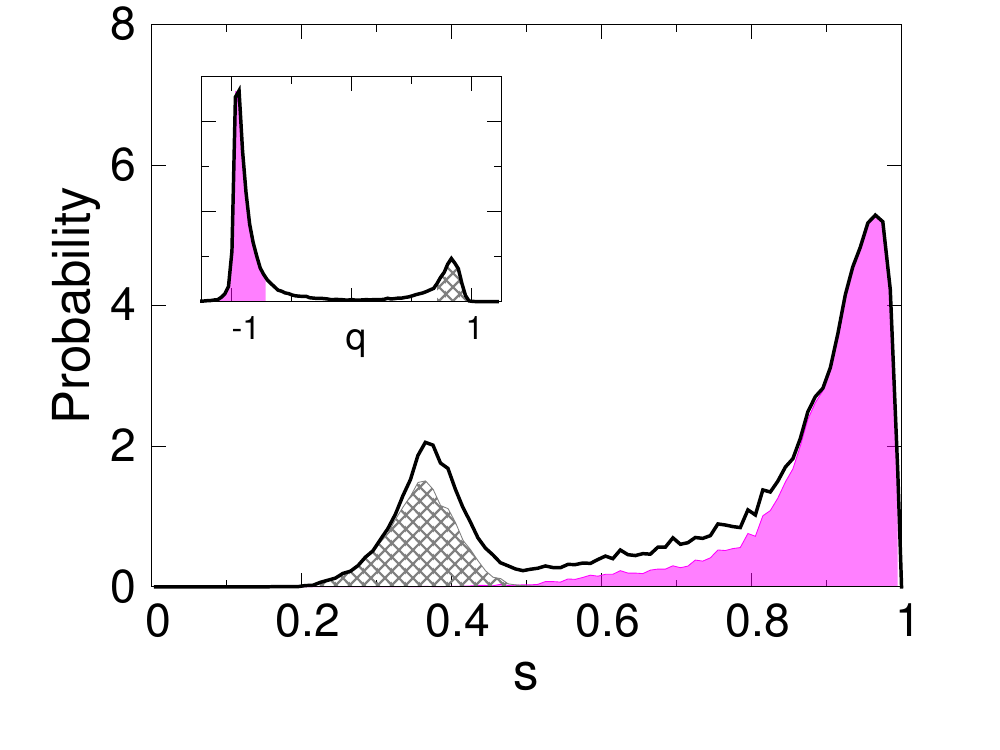}

\caption{(Color online) $(19/2)^{6}$. The distribution of the relative transition
strength $s$ for the collective realizations (shaded area in Fig.
\ref{fig:jl19N6}). The quadrupole moments shown in the inset are
separated into prolate ($q>0.7$) and oblate ($q<-0.7$) shapes. The
resulting distributions are shaded with a pattern and a uniform color,
respectively. The fraction of oblate cases is 5.2\% and the fraction
of prolate cases is 1.3\% relative to the total number of random realizations. }
\label{fig:jl19N6_COL}
\end{figure}

The collective realizations are further analyzed in Fig. \ref{fig:jl19N6_COL}
where the distribution of the relative transition strength $s$ is
shown. In Fig. \ref{fig:jl19N6_COL} the oblate $(q<-0.7)$ and prolate
$(q>0.7)$ cases are shaded with different patterns. The relative
transition rate $s$ for the oblate samples is close to the maximum
possible $s=1.$ Thus, for these realizations the ground state band
structure is similar to that of the QQ Hamiltonian. The data on the
QQ Hamiltonian for our models is summarized in Sec.\ref{sec:Summary}.
For prolate systems the distribution of the relative transition strength
peaks around $s=0.37$. 

\begin{figure}
\includegraphics[width=3.4in]{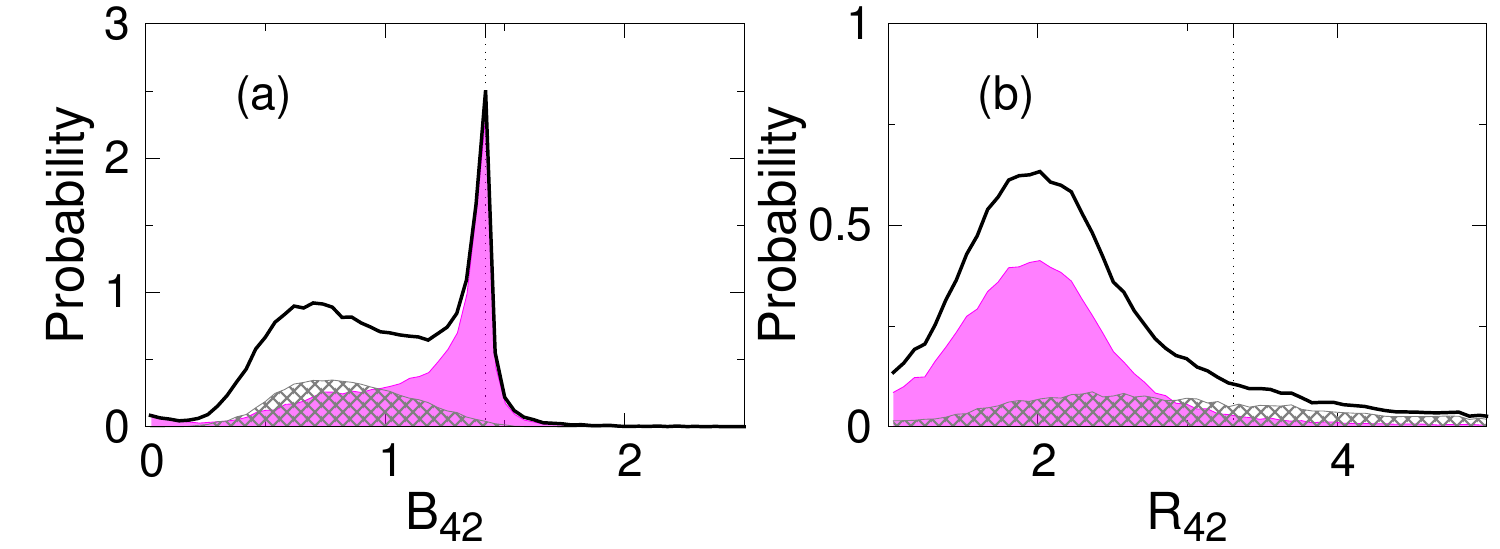}

\caption{(Color online) $(19/2)^{6}$. (a) The distribution of the deexcitation
ratio $B_{42}$ defined in Eq. (\ref{eq:DX}). (b) The distribution
of the excitation energy ratio $R_{42}$ defined in Eq. (\ref{eq:RR}).
The distributions are comprised of 13.6\% of realizations that have
the $0_{gs},\,2_{1},\,4_{1}$ sequence with $b>0.7$, the $2_{1}$
state is not higher than the fourth excited state, and $E(4_{1})>E(2_{1})$.
The prolate cases and oblate cases, that appear in the ensemble with
probabilities 3.3\% and 7.1\% respectively, are shaded with the same
patterns as in Fig. \ref{fig:jl19N6_COL}. The values of $B_{42}$
and $R_{42}$ for the QQ Hamiltonian listed in Tab. \ref{tab:QQ}
are marked with the vertical grid lines.}
\label{fig:jl19N6_COL024}
\end{figure}

In Fig. \ref{fig:jl19N6_COL024} we focus on distributions of the
deexcitation ratio $B_{42}$ and the energy ratio $R_{24}$ defined
in Eqs. \ref{eq:DX} and \ref{eq:RR}. We use the same shading for
prolate and oblate realizations as in Fig. \ref{fig:jl19N6_COL},
but slightly modify our selection of samples. We chose the collective
realizations that have states $2_{1}$ and $4_{1},$ with $2_{1}$
being not higher than the forth excited state and $4_{1}$ state being
above it. 

The collective oblate realizations comprise a peak in the distribution
of $B_{42}$ in Fig. \ref{fig:jl19N6_COL024}(a) around the rotational
limit of $B_{42}=1.4$. For prolate realizations the distribution
peaks near $B_{42}=0.8$ and has an extended shoulder. It is likely
that the rotational structure is fragmented in instances with a weak
prolate deformation. Here the lower value of $s$ seen in Fig. \ref{fig:jl19N6_COL}
is used to suggest a weak deformation. Therefore, the $4_{1}$ state
is not purely rotational. 

The distribution of the ratio of the excitation energies $R_{42}$
in Fig. \ref{fig:jl19N6_COL024}(b) seems to contradict the rotational
limit. For most of the collective realizations the values of the ratio
fall between the pairing limit of $1$ and the vibrational limit of
$2$, while in the rotational limit the ratio of $3.3$ is expected.
This discrepancy has been reconciled in Ref. \cite{Johnson:1998}
with the observation that the rotational ordering emerges for the
ensemble-averaged excitation energies. The same conclusion is expected
from the geometrical chaoticity arguments \cite{Mulhall:2000}. Excitation
energies are sensitive to non-collective features, this leads to large
fluctuations of $R_{42}$. The experimental observations of realistic
nuclei also show that when the quadrupole transition rates follow
the rotational systematics, the excitation energy spectrum can deviate
from rotational; on occasions, the spectrum is closer to the vibrational
limit \cite{Casten:2000}. The coexistence of both prolate and oblate
configurations in this $(19/2)^{6}$ system could be another reason
for the distortion in the energy spectrum. Within Elliot's SU(3) model
analogous mixing of group representations was investigated in Ref.
\cite{Thiamova:2006}. 

\subsection{Triaxiality\label{sub:Triaxiality}}

The triaxiality is marked by the presence of low-lying levels $2_{2},\:3_{1},\:4_{2},\:5_{1}$.
The excitation energies are subject to equalities $E(2_{1})+E(2_{2})=E(3_{1})$
and $4E(2_{1})+E(2_{2})=E(5_{1})$. It is remarkable that these relations
appear to be well satisfied by the spectrum of the QQ Hamiltonian
in the $(19/2)^{6}$ configuration, for which $R_{2_{1}3_{1}}+R_{2_{2}3_{1}}=1.005$
and $4R_{2_{1}5_{1}}+R_{2_{2}5_{1}}=1.026.$ Here $R_{JJ'}=E(J)/E(J')$
denotes the ratio of excitation energies. The rigid rotor Hamiltonian,
defined by three moments of inertia, is responsible for these correlations
in the spectrum. 

In this work we examine two low-lying $2_{1}$ and $2_{2}$ states.
These are the only states with spin 2 in the triaxial rotor model,
they are mixed configurations of $K=0$ and $K=2$. We use angle $\Gamma$
to express the level of the $K$-mixing. This angle is determined
by the three reciprocal moments of inertia $A_{i}$, $i=1,2,3$ in
the rotor Hamiltonian\[
\tan2\Gamma=\frac{\sqrt{3}(A_{1}-A_{2})}{A_{1}+A_{2}-2A_{3}}.\]
The ratio of the excitation energies of the $2_{1}$ and $2_{2}$
states is another parameter of the rotor Hamiltonian. It is convenient
to express this ratio $R_{2_{1}2_{2}}$ in terms of the angle $\gamma_{DF}$
defined using the Davydov-Filippov model of irrotational flow \cite{Davydov:1958}
as\begin{equation}
\text{\ensuremath{\sin}}^{2}(3\gamma_{DF})=\frac{9}{2}\frac{R_{2_{1}2_{2}}}{\left(1+R_{2_{1}2_{2}}\right)^{2}}.\label{eq:trixgammaDF}\end{equation}
In our example that follows, the triaxiality is small and $\gamma_{DF}^{2}\approx0.5R_{2_{1}2_{2}}$.
Thus, the rotor Hamiltonian, given by the three moments of inertia,
can be equivalently described by an overall energy scale, the $K$-mixing
angle $\Gamma$, and the angle $\gamma_{DF}$. 

The quadrupole shape is parametrized by the Hill-Wheeler parameters
$\beta$ and $\gamma$ which define the quadrupole operator ${\cal M}_{2\mu}$.
The relation between the parameters of the rotor Hamiltonian and the
intrinsic shape is model-dependent. The irrotational-flow moments
of inertia discussed in Ref. \cite{Davydov:1958} result in $\gamma_{DF}=\gamma$
and $\Gamma=\left\{ {\rm arccot}\left[3\cot(3\gamma)\right]-\gamma\right\} /2$;
the latter implies $\Gamma\ll\gamma$ for small triaxiality. A rather
different result follows from the rigid-body moments of inertia. 

We determine $\Gamma,$ $\gamma_{DF},$ and $\gamma$ independently
from the spectroscopic observables. The parameter $\gamma_{DF}$ is
obtained from the energy spectrum, Eq. (\ref{eq:trixgammaDF}). Following
Ref. \cite{Allmond:2008} one can view the sum rules \[
b(E2,0_{gs}\rightarrow2_{1})+b(E2,0_{gs}\rightarrow2_{2})=1\]
and \[
\frac{7}{2}b(E2,2_{1}\rightarrow2_{2})+q^{2}(2_{1})=1\]
for the $J=2$ two-state model as the Pythagorean theorem for amplitudes.
The angles in the corresponding right-angled triangles are $\gamma-\Gamma$
and $\gamma+2\Gamma,$ therefore \begin{equation}
\tan^{2}(\gamma-\Gamma)=\frac{B(E2,0\rightarrow2_{2})}{B(E2,0\rightarrow2_{1})},\label{eq:trixgamma}\end{equation}

\begin{equation}
\tan^{2}(\gamma+2\Gamma)=\frac{2B(E2,2_{1}\rightarrow2_{2})}{7{\cal Q}^{2}(2_{1})}.\label{eq:trixgammaQ}\end{equation}
These equations allow one to determine the triaxiality $\gamma$ and
the $K$-mixing angle $\Gamma$. 

All three angles $\gamma,$ $\Gamma,$ and $\gamma_{DF}$ are small
in our models with the QQ Hamiltonian, see discussion in Sec. \ref{sec:Summary}.
Correspondingly, in the TBRE the effects of triaxiality are weak but
detectable. 

\begin{figure}
\includegraphics[width=3in]{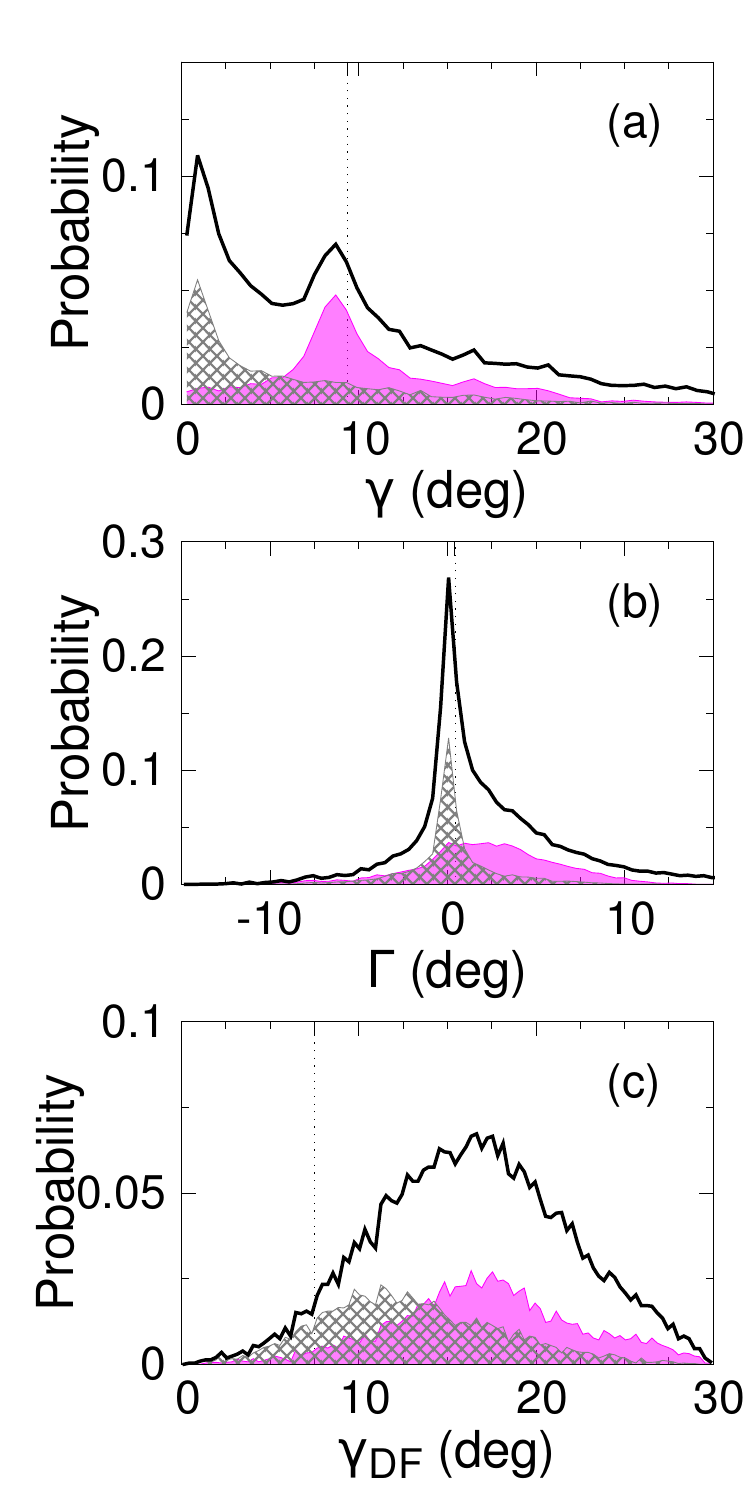}

\caption{(Color online) $(19/2)^{6}$. (a) The distribution of the triaxiality
angle $\gamma$ . (b) The distribution of the $K$-mixing angle $\Gamma$.
(c) The distribution of the triaxiality angle $\gamma_{DF}$ from
the Davydov-Filippov model. The angles are obtained from Eqs. (\ref{eq:trixgamma}),
(\ref{eq:trixgammaQ}) and (\ref{eq:trixgammaDF}). We select realizations
with two states of spin 2 in the spectrum and require $b>0.7$ and
$q(2_{1})\approx-q(2_{2})$; 18.3\% of realizations satisfy this set
of restrictions. The realizations with prolate and oblate shapes are
shaded with the same patterns as in Figs. \ref{fig:jl19N6_COL} and
\ref{fig:jl19N6_COL024}. Vertical grid lines indicate the triaxiality
parameters calculated from the QQ Hamiltonian, which are: $\gamma=9.79,$
$\Gamma=0.73,$ and $\gamma_{DF}=7.52.$}
\label{fig:jl19N6_trix}
\end{figure}

For our studies of triaxiality presented in Fig. \ref{fig:jl19N6_trix}
we use the $(19/2)^{6}$ model. We recall that in the triaxial rotor
model there is a second $2_{2}$ state with ${\cal Q}(2_{1})=-{\cal Q}(2_{2}).$
Thus, we select collective realizations with $0_{gs}$ and $2_{1}$,
and in addition to that we require that in the entire spectrum there
is a $2_{2}$ state for which the equality ${\cal Q}(2_{2})=-{\cal Q}(2_{1})$
holds within 20\% of accuracy. In collective realizations of rotational
type the magnitude of the $Q(2_{2})$ is large as compared to the
quadrupole moments of other many-body states. This simplifies the
identification of the $2_{2}$ state. We find that practically for
all collective realizations this second $2_{2}$ state exists. Indeed,
from the total number of random realizations a large fraction, 18.3\%,
satisfy all of the mentioned triaxiality conditions. In Figs. \ref{fig:jl19N6_trix}(a),
\ref{fig:jl19N6_trix}(b), and \ref{fig:jl19N6_trix}(c) we show the
distributions of the triaxiality angle $\gamma$, $K$-mixing angle
$\Gamma,$ and $\gamma_{DF}$, respectively. We use the same shading
as in Fig. \ref{fig:jl19N6_COL024} to separate prolate and oblate
shapes.

In the $(19/2)^{6}$ model one often finds collective realizations
with oblate intrinsic deformation and $s\approx1$, these realizations
are triaxial with $\gamma\approx9^{\circ}$, Fig. \ref{fig:jl19N6_trix}(a).
This result, as well as $\Gamma\approx0^{\circ}$ in Fig. \ref{fig:jl19N6_trix}(b),
is consistent with that of the QQ Hamiltonian. The less frequent prolate
cases are nearly axially symmetric. 

In the TBRE the angle $\gamma_{DF}$, Fig. \ref{fig:jl19N6_trix}(c),
appears on average to be higher than the corresponding angle in the
QQ Hamiltonian. The peak in the $\gamma_{DF}$ distribution is also
higher than the peak in the $\gamma$ distribution, compare Figs.
\ref{fig:jl19N6_trix}(a) and \ref{fig:jl19N6_trix}(c). (We remind
that $\gamma_{DF}=\gamma$ in the irrotational flow model.) Nevertheless,
no conclusions can be made from these two discrepancies. We believe
that the excitation energies could be influenced significantly by
non-collective features. The situation may be similar to the one in
Fig. \ref{fig:qqj19N8}(b), where $4_{1}$ state is lower than expected
for the rotor. Similarly, if the $2_{2}$ state is lowered the resulting
$\gamma_{DF}$ is larger. In both cases the lowering is relative to
the excitation energy of the $2_{1}$ state.

\subsection{Higher multipole moments\label{sub:Higher-multipole-collectivity}}

\begin{figure}
\includegraphics[width=3.4in]{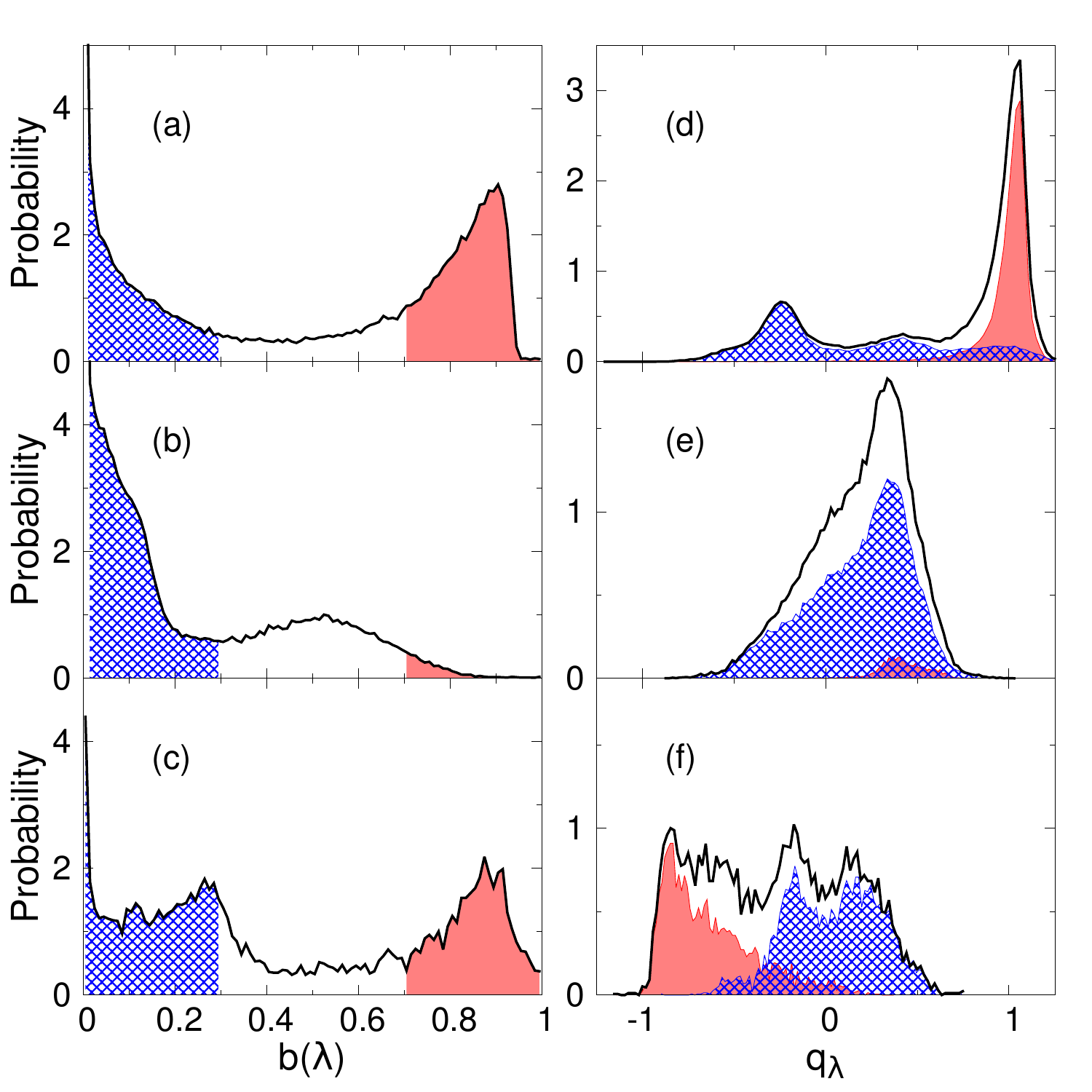} \caption{(Color online) $(19/2)^{6}$. The distributions of the fractional
collectivity $b(\lambda)$ are shown in panels (a), (b), and (c).
The distributions of the intrinsic multipole moments $q_{\lambda}$
are shown in panels (d), (e), and (f). The plots are organized in
three rows corresponding to multipolarities with $\lambda=4,\:6\text{ and }8$.
Here we include realizations where, in addition to the $0_{gs}$ state,
the first excited state is either $4_{\boldsymbol{1}},$ or $6_{\boldsymbol{1}},$
or $8_{\boldsymbol{1}}.$ The shaded areas correspond to collective
and non-collective modes with $b(\lambda)>0.7$ and $b(\lambda)<0.3$
respectively. We use the same patterns as in Fig. \ref{fig:jl19N6}.}
\label{fig:multipole19N6}
\end{figure}

It is known that in the TBRE the probability to find a $0_{gs}$ state
followed by either one of the states $2_{\boldsymbol{1}},$ $4_{\boldsymbol{1}},$
$6_{\boldsymbol{1}},$ or $8_{\boldsymbol{1}}$ is disproportionally
large as compared to what is statistically expected. For the $(19/2)^{6}$
model the corresponding probabilities are 10.4\%, 17.3\%, 11.9\% and
1.8\%. In an attempt to understand this, we repeat the previous study
but target the collective realizations of multipolarity $\lambda=4,\,6$,
and 8. For realizations with the $0{}_{gs}$ state and with the first
excited state of spin $\lambda$ in Fig. \ref{fig:multipole19N6},
we consider the fractional collectivity $b(\lambda)\equiv b(E\lambda,0_{gs}\rightarrow\mbox{\ensuremath{\lambda}}_{1})$
and the multipole moment $q_{\lambda}\equiv q_{\lambda}(\lambda_{1})$.
Fig. \ref{fig:multipole19N6} shows evidences for intrinsic shapes
with deformations of higher multipolarities. In particular, for $\lambda=4$
and 8 there is a sizable number of collective realizations where $b(\lambda)>0.7$.
These realizations are shaded (in red). The corresponding distributions
of the multipole moments in Figs. \ref{fig:multipole19N6}(d) and
\ref{fig:multipole19N6}(f) have peaks which are centered at non-zero
values of $q_{\lambda}$. The $\lambda=6$ shape collectivity is nearly
absent in the $(19/2)^{6}$ system: the realizations are mostly non-collective,
$b(6)<0.7$ (shaded in blue), and the corresponding moment has a peak
centered near zero. 

Investigations of other single $j$ systems show presence of multipole
collectivities with $\lambda=2,4,6,\text{ and }8$. Generally, the
collectivities corresponding to the intrinsic quadrupole shape are
the most pronounced ones, however there are signatures of realizations
with shapes of higher multipole deformation. The existence of such
geometric structures may be related to the symmetries discussed in
Ref. \cite{Huu-Tai:2002}

\subsection{Multipole structure of the Hamiltonian}

\begin{figure}
\includegraphics[width=3.4in]{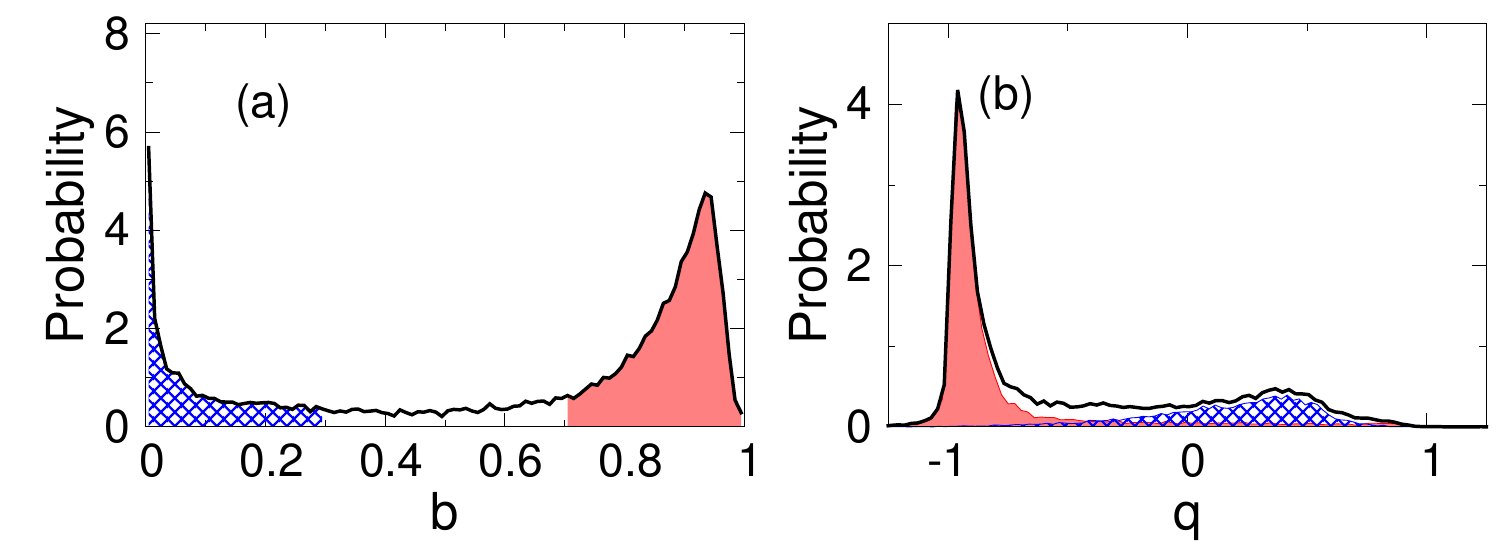}\caption{(Color online) $(19/2)^{8}$. The same figure as Fig. \ref{fig:jl19N6}
but for the 8-particle system. (a) The distribution of the fractional
collectivity $b$. (b) The distribution of the intrinsic quadrupole
moment $q$. The histogram is comprised of 7.5\% of random spectra
with $0_{gs}$ and $2_{\boldsymbol{1}}$ states. Shaded areas correspond
to 4.6\% of collective realizations ($b>0.7$) and 1.9\% of non-collective
realizations ($b<0.7$). This figure is analogous to Figs. \ref{fig:jl19N6}
and \ref{fig:multipole19N6}, and the same shading is used in these
figures.}
\label{fig:qqj19N8}
\end{figure}

\begin{figure}
\includegraphics[width=3.4in]{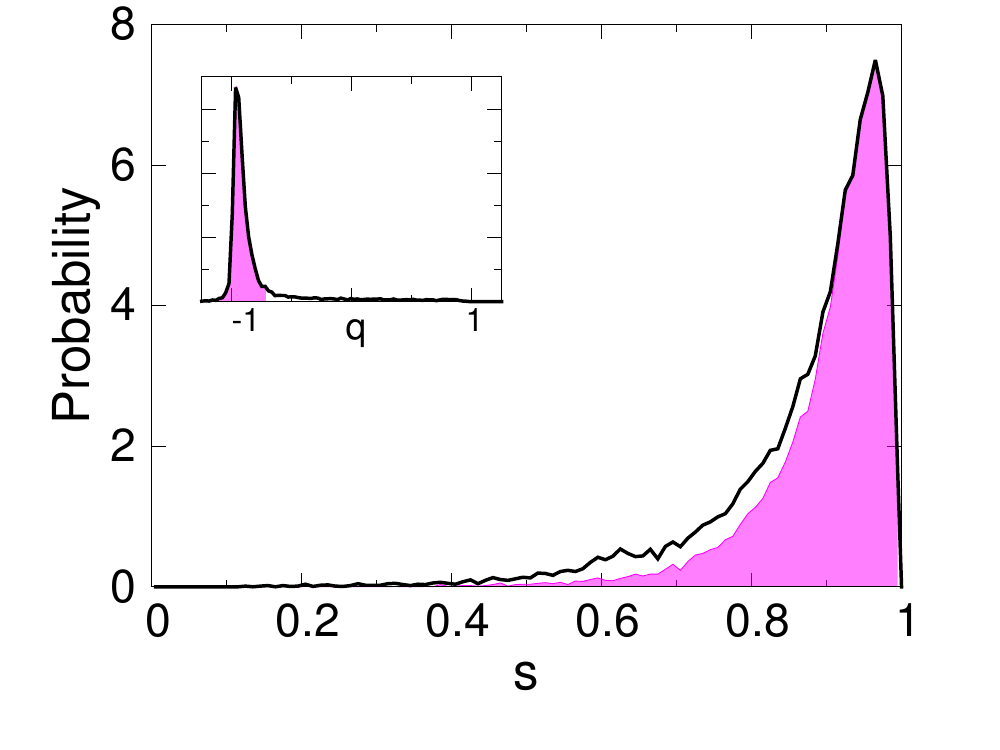}\caption{(Color online) $(19/2)^{8}$. The same figure as Fig. \ref{fig:jl19N6_COL}
but for the 8-particle system. The distribution of the relative transition
strength rule $s$ for the collective realizations. This figure is
analogous to Fig. \ref{fig:jl19N6_COL}, and the same shading is used
as in Figs. \ref{fig:jl19N6_COL} and \ref{fig:jl19N6_COL024}, however
only oblate shapes $(q<-0.7)$ are seen. }
\label{fig:COLqqj19N8}
\end{figure}

In this subsection we discuss the multipole structure of the two-body
Hamiltonian in the single $j$ level model. For this purpose we use
a larger system of $8$ nucleons in the same $j=19/2$ model space,
i.e. the $(19/2)^{8}$ model. The distributions of the fractional
collectivity, the quadrupole moment, and the relative transition strength
shown in Figs. \ref{fig:qqj19N8} and \ref{fig:COLqqj19N8} are similar
to the distributions observed in the $(19/2)^{6}$ model in Figs.
\ref{fig:jl19N6} and \ref{fig:jl19N6_COL}. The main difference between
the models is that, in contrast to Fig. \ref{fig:jl19N6}(b), only
oblate ground state configurations are present in Fig. \ref{fig:qqj19N8}(b). 

The collectivities observed in the single $j$ studies are deeply
rooted in the underlying geometric structure of the Hamiltonian. To
focus on this relation we express the two-body Hamiltonian in the
particle-hole channel in terms of the multipole operators

\begin{equation}
H=\sum_{\mathcal{K}}\tilde{V}_{\mathcal{K}}\sum_{\varkappa}{\cal M}_{\mathcal{K}\varkappa}^{\dagger}{\cal M}_{\mathcal{K}\varkappa}.\label{eq:H_int_ph}\end{equation}
The interaction parameters $\tilde{V}_{\mathcal{K}}$ in the particle-hole
channel are determined from those in the particle-particle channel
$V_{L}$ via Pandya transformation 

\begin{equation}
\tilde{V}_{\mathcal{K}}=\sum_{L}(2L+1)\chi_{L}^{\mathcal{K}}\, V_{L}.\label{eq:PP2PH}\end{equation}
The transformation coefficients \[
\chi_{L}^{\mathcal{K}}=\left\{ \begin{array}{ccc}
j & j & \mathcal{K}\\
j & j & L\end{array}\right\} ,\]
are given by the six-$j$ recoupling coefficients. 

On a single $j$ level only even values of the particle-pair angular
momenta $L$ are allowed by the Fermi statistics. Thus, there are
$j+1/2$ interaction parameters $V_{L}$ in Eq. (\ref{eq:PP2PH}).
In studies of the TBRE a set of these parameters can be viewed as
a random vector in the $j+1/2$ dimensional space. There is no such
a restriction on the particle-hole momentum $\mathcal{K}$. Thus,
the inverse transformation \begin{equation}
V_{L}=\sum_{\mathcal{K}}(2\mathcal{K}+1)\chi_{\mathcal{K}}^{L}\,\tilde{V}_{\mathcal{K}}\label{eq:VLK}\end{equation}
may produce some unphysical $V_{L}$ with odd values of $L.$ Such
Pauli-forbidden terms in the Hamiltonian do not generate any dynamics.
Therefore the $2j+1$ parameters $\tilde{V}_{\mathcal{K}}$ contain
passive components which can be removed making $\tilde{V}_{\mathcal{K}}$
linearly dependent \cite{Volya:2002PRC}. 

The interaction terms that correspond to the multipoles with momentum
$\mathcal{K}=0$ and $\mathcal{K}=1$ are constants of motion \cite{Zelevinsky:2000}.
The $\mathcal{K}=0$ term in Eq. (\ref{eq:H_int_ph}), describes the
nucleon-nucleon interaction that is the same for all angular momentum
channels, $V_{L}=\chi_{L}^{0}={\rm const}$, as follows from Eq. (\ref{eq:VLK}).
The resulting monopole Hamiltonian is proportional to the number of
particle-pairs in a system. This Hamiltonian has no dynamical effect.
Thus, there is no change in results if one constrains the TBRE by
projecting out the monopole $\mathcal{K}=0$ component as follows
\begin{equation}
V_{L}\rightarrow V_{L}-\chi^{0}\frac{\sum_{L'}(2L'+1)\chi_{L'}^{0}\, V_{L'}}{\sum_{L'}(2L'+1)\left(\chi_{L'}^{0}\right)^{2}\,}.\label{eq:projection}\end{equation}
This effectively reduces the number of independent parameters $V_{L}$. 

In a single $j$ model space the $\mathcal{K}=1$ multipoles are proportional
to the angular momentum operators ${\cal M}_{1\varkappa}\sim J_{\varkappa}$.
Therefore the $\mathcal{K}=1$ interaction leads to a rotational $E(J)\sim J(J+1)$
spectrum with $\tilde{V}_{\mathcal{K}=1}$ determining the moment
of inertia. In the particle-particle channel, the $J^{2}$ operator
is obtained with $V_{L}=\chi_{L}^{1}\sim{\rm const}+L(L+1)$. Consistently,
it was argued in Refs. \cite{Zelevinsky:2004,Volya:2008} that those
interactions that lead to the positive moment of inertia are likely
to result in the $J_{gs}=0$. The exact $J^{2}$ operator component
in the interaction can be removed by orthogonalization to $\chi_{L}^{1}$
following the procedure in Eq. (\ref{eq:projection}). 

\begin{figure}
\includegraphics[width=3.4in]{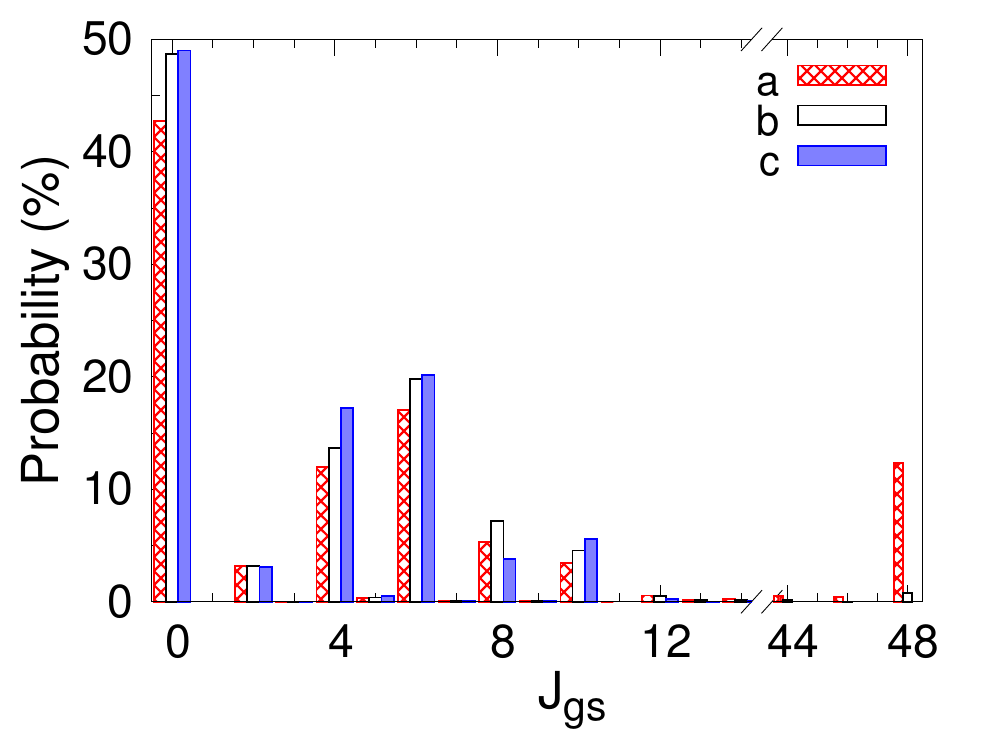}\caption{(Color online) $(19/2)^{8}$. Probabilities to observe a certain ground
state spin $J_{gs}$ for three random ensembles: (a) the TBRE, (b)
the TBRE without a $J^{2}$ term (the $\mathcal{K}=1$ term in Eq.
(\ref{eq:H_int_ph}) is removed), and (c) the TBRE without both, $J^{2}$
and QQ terms (the $\mathcal{K}=1$ and $\mathcal{K}=2$ terms in Eq.
(\ref{eq:H_int_ph}) are removed).}
\label{fig:j19N8gsspin}
\end{figure}

\begin{figure}
\includegraphics[width=3.4in]{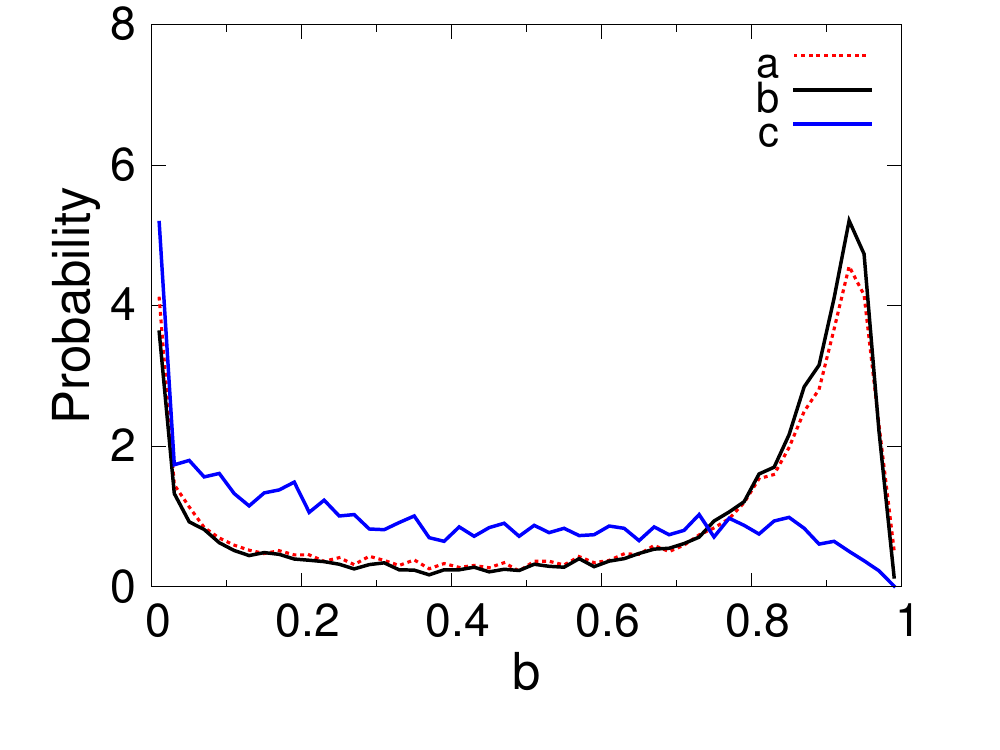}\caption{(Color online) $(19/2)^{8}$: The distribution of the fractional collectivity
$b$ for the same three random ensembles as in Fig. \ref{fig:j19N8gsspin}.
Namely: (a) the traditional TBRE, (b) the TBRE without a $J^{2}$
term, and (c) the TBRE without both, $J^{2}$ and QQ terms. We select
realizations with the $0_{gs}$ state followed by the first excited
state $2_{\boldsymbol{1}}$. The fraction of such cases for ensembles
(a), (b), and (c) is 7.6\%, 8.2\%, and 4.7\% respectively. }
\label{fig:multi_j19N8_JJ}
\end{figure}

The changes in dynamics are no longer trivial when the quadrupole
$\mathcal{K}=2$ component in the interaction is modified. The role
of different multipoles in the TBRE is studied in Fig. \ref{fig:j19N8gsspin}
and \ref{fig:multi_j19N8_JJ} where we remove different $\mathcal{K}$
components from the interaction Hamiltonian in Eq. (\ref{eq:H_int_ph})
using the Graham-Schmidt projection procedure. In the particle-particle
channel the projection of pairing interaction $V_{L}=\delta_{L,0}$
has been extensively discussed in Ref. \cite{Zelevinsky:2004}. The
removal of pairing does not lead to any significant qualitative change,
we thus forgo this topic in what follows. 

The probability to observe a certain ground state spin in the $(19/2)^{8}$
system is shown in Fig. \ref{fig:j19N8gsspin}. Three cases of random
ensembles are reviewed: (a) the traditional TBRE where all $j+1/2$
interaction parameters $V_{L}$ are random Gaussian variables, (b)
the case where $\mathcal{K}=1$ term is removed, and (c) the ensemble
where \textit{$\mathcal{K}=1$ }and\textit{ $\mathcal{K}=2$} multipole
components are removed from the Hamiltonian. While the wave functions
in ensembles (a) and (b) are identical, the ground state spin distributions
are different. The role of the $J^{2}$ moment-of-inertial-like term
has been discussed before in Refs. \cite{Mulhall:2000,Zelevinsky:2004,Volya:2008};
it appears to be fully responsible for the cases with maximum possible
spin. As seen in Fig. \ref{fig:j19N8gsspin}, the states with the
maximum spin almost never occur as ground states in ensembles (b)
and (c) where the $J^{2}$ interaction term ($\mathcal{K}=1$) is
removed. 

The ensembles (b) and (c) shown in Fig. \ref{fig:j19N8gsspin} appears
to have similar ground state spin distributions but the behavior of
the fractional collectivity is different. In Fig. \ref{fig:multi_j19N8_JJ}
for all three ensembles we show the distribution of the fractional
collectivity of the transition between the $0_{gs}$ and $2_{\boldsymbol{1}}$
states. It is evident that the quadrupole collectivity disappears
once the quadrupole component in the interaction is removed. Thus,
we conclude that the quadrupole-quadrupole component in the interaction
generates the corresponding deformation and is responsible for the
rotational behavior observed.

\section{Models beyond single $j$ \label{sec:two_levels}}

\begin{figure}
\includegraphics[width=3.4in]{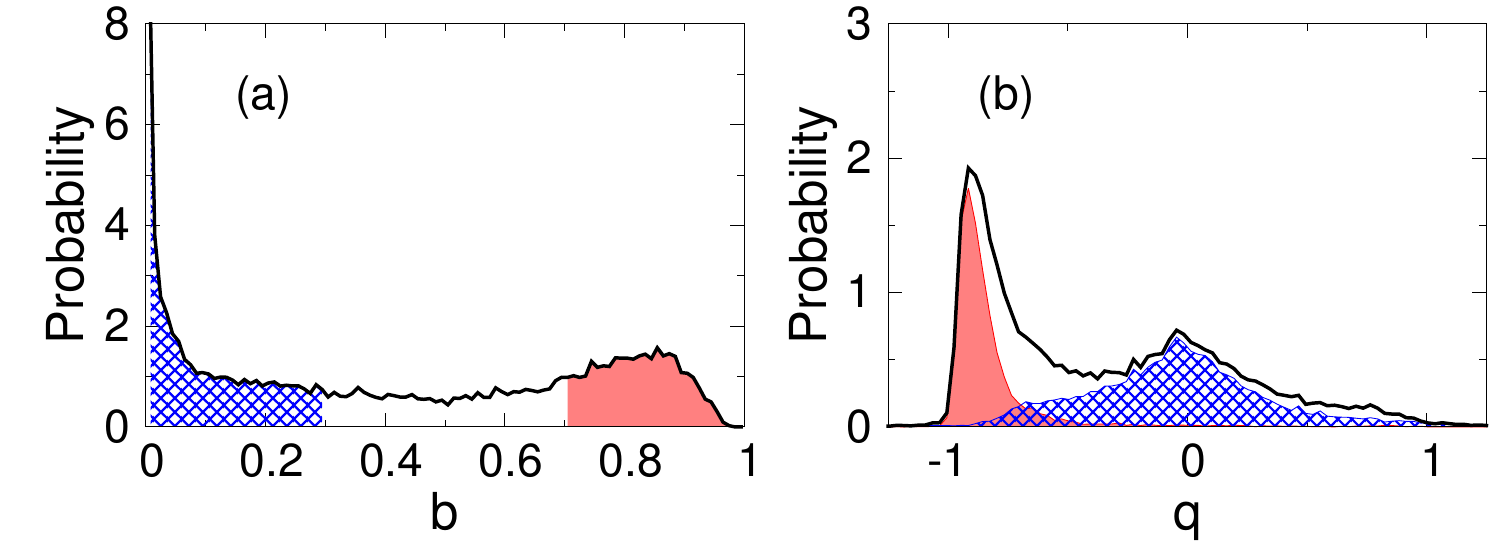}

\caption{(Color online) $(13/2^{+},13/2^{+})^{6}$. (a) The distribution of
the fractional collectivity $b$. (b) The distribution of the intrinsic
quadrupole moment $q$. The $4.1\%$ of samples have the $0_{gs},2_{\boldsymbol{1}}$
sequence. Shaded areas correspond to 1.2\% of collective realization
and to 1.8\% of non-collective realizations. This figure is analogous
to Figs. \ref{fig:jl19N6}, \ref{fig:multipole19N6}, and \ref{fig:qqj19N8},
and the same shading is used in these figures. }

\label{fig:ppvj13N6}
\end{figure}

In this section we expand the scope of our models and consider systems
with two single-particle levels. The richer geometry allows one to
study the effects of particle-hole conjugation, different structures
of the multipole operators, and the role of parity. The distributions
of the fractional collectivity and of the quadrupole moment are shown
in Figs. \ref{fig:ppvj13N6} for the $(13/2^{+},13/2^{+})^{6}$ system.
Here the model space is comprised of two levels with $j_{1}=j_{2}=13/2$.
Both single-particle levels have the same positive parity, so that
the effective spherical Hartree-Fock mean-field Hamiltonian can contain
terms of a mixed structure such as $a_{j_{1}}^{\dagger}a_{j_{2}}.$
These terms are scalars for $j_{1}=j_{2}.$ There is some arbitrariness
in the choice of the single-particle matrix elements of the multipole
operator ${\cal M}_{2\mu}$ which depend on the radial overlap of
the operator $r^{2}$. We choose the radial overlap to be diagonal
$\langle j_{1}|r^{2}|j_{1}\rangle=\langle j_{2}|r^{2}|j_{2}\rangle$
and $\langle j_{1}|r^{2}|j_{2}\rangle=0$; other possibility with
$\langle j_{1}|r^{2}|j_{1}\rangle=\langle j_{2}|r^{2}|j_{2}\rangle=\langle j_{1}|r^{2}|j_{2}\rangle$
has been explored and led to no substantial difference. 

\begin{figure}
\includegraphics[width=3.4in]{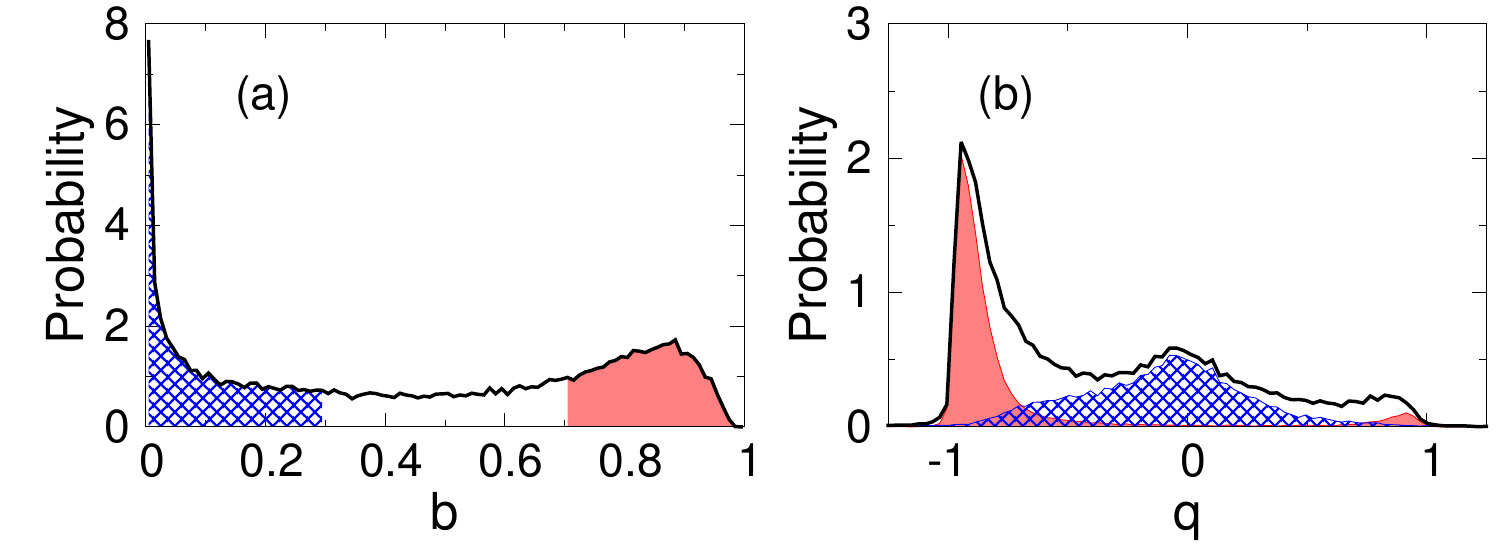}

\caption{(Color online) $(13/2^{+},13/2^{-})^{6}$. (a) The distribution of
the fractional collectivity $b$. (b) The distribution of the intrinsic
quadrupole moment $q$. The $6.9\%$ of samples have the $0_{gs}$
state and the first state $2_{1}$, both of positive parity. The $2.4\%$
of collective and $2.6\%$ of non-collective realizations are shaded
with patterns. This figure is analogous to Figs. \ref{fig:jl19N6},
\ref{fig:multipole19N6}, \ref{fig:qqj19N8}, and \ref{fig:ppvj13N6},
and the same shading is used in these figures. }

\label{fig:vj13N6}
\end{figure}

A structurally different $(13/2^{+},13/2^{-})^{6}$ model is examined
in Fig. \ref{fig:vj13N6} where two levels of equal spin and different
parity are considered. In this case the matrix elements of the Hamiltonian
are restricted by parity. The same structure of the quadrupole operator
is used. The model space of this kind has been explored in Ref. \cite{Zelevinsky:2008}
because it is the simplest model space that allows for quadrupole
and octupole modes. The prevalence of the positive parity ground states
is remarkable in this model. The ground state is most likely to have
spin-parity $0^{+}$, $4^{+},$ or $24^{+}$ with 35\%, 19\%, and
14\% probability, respectively. In contrast, $0^{-}$, the most probable
negative parity ground state, happens only in 3\% of realizations.
For the $(13/2^{+},13/2^{-})^{6}$ model the number of many-body states
is the same for both parities, 8,212 each. 

For both $(13/2^{+},13/2^{+})^{6}$ and $(13/2^{+},13/2^{-})^{6}$
models, the results related to the quadrupole collectivity are almost
identical, see Figs. \ref{fig:ppvj13N6} and \ref{fig:vj13N6}. Moreover,
these results are similar to those for the single $j$ level models,
compare to Figs. \ref{fig:jl19N6} and \ref{fig:qqj19N8}. The major
features in the distributions of $b$ and $q$ persist despite a bigger
number of random parameters defining the Hamiltonians, more complex
geometry of the two-level models, and a more chaotic resulting dynamics.
There is a peak in the distribution of the fractional collectivity
$b$ near 1 indicating a sizable number of collective cases. The distribution
of the quadrupole moment for the collective realizations (shaded in
red) has a well-defined peak on the oblate side. The non-collective
realizations appear to have quadrupole moment distribution centered
at zero (shaded in blue). 

\begin{figure}
\includegraphics[width=3.4in]{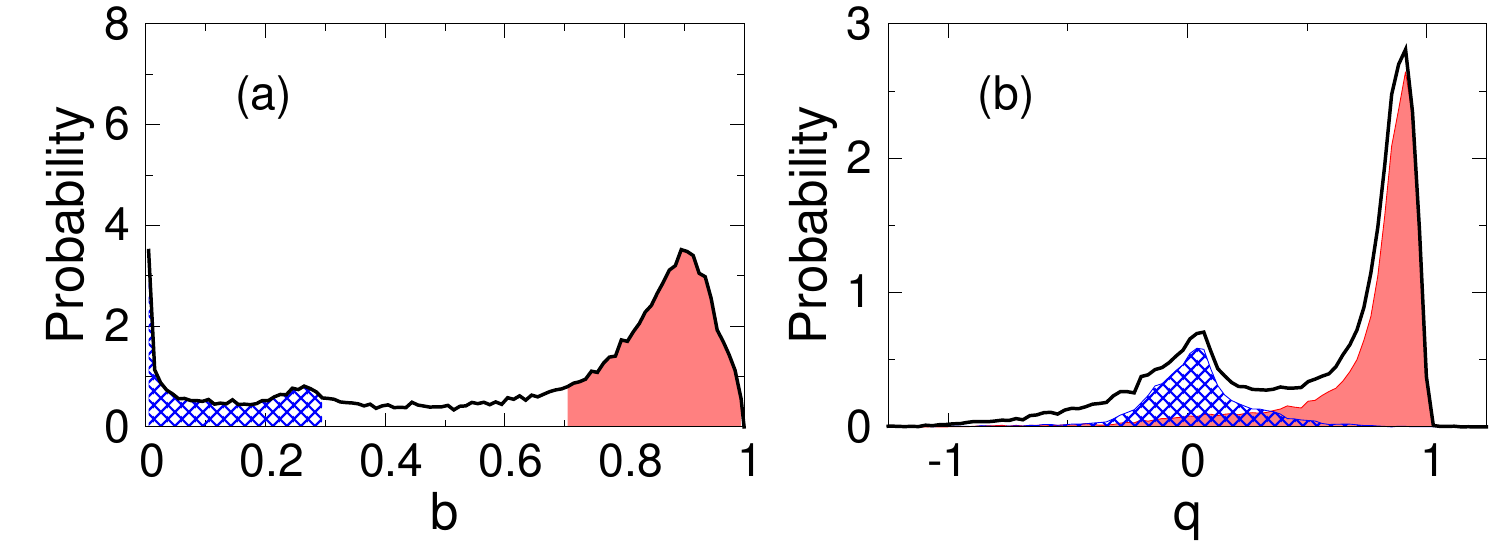}\caption{(Color online) $(13/2^{+},13/2^{+})^{22}$. (a) The distribution of
the fractional collectivity $b$. (b) The distribution of the intrinsic
quadrupole moment $q$. The system is particle-hole conjugated to
that in Fig. \ref{fig:ppvj13N6}. The percentage of samples with the
$0_{gs},2_{\boldsymbol{1}}$ sequence is $8.1\%$ which includes $4.8\%$
of collective and $1.7\%$ of non-collective. This figure is analogous
to Figs. \ref{fig:jl19N6}, \ref{fig:multipole19N6}, \ref{fig:qqj19N8},
\ref{fig:ppvj13N6}, and \ref{fig:vj13N6}, and the same shading is
used in these figures. }

\label{fig:ppvj13N22}
\end{figure}

For systems with exact particle-hole symmetry the quadrupole moment
for particles is equal in magnitude and opposite in sign to that of
holes. Moreover, properties such as excitation energies, spins of
states, and transition rates, are exactly equal for particle-hole
conjugated systems. The particle-to-hole transformation for any two-body
Hamiltonian amounts to the same Hamiltonian for holes but with an
additional one-body term. Thus, the symmetry is not exact in a two-level
model space. Nevertheless in the TBRE, where two-body matrix elements
are selected symmetrically about zero, the one-body term averages
to zero. Therefore the results in Figs. \ref{fig:ppvj13N22} and \ref{fig:ppvj13N6}
for particle-hole conjugates systems $(13/2^{+},13/2^{+})^{22}$ and
$(13/2^{+},13/2^{+})^{6}$ are nearly symmetric. The main difference
is that the Hamiltonian for holes contains random single-particle
energies which leads to a different fraction of collective realizations
in the ensembles. 

\section{Realistic Valence Space\label{sec:Realistic-Model-Space}}

The schematic models discussed in the previous sections all seem to
possess the rotational low-lying spectrum which is an evidence of
the intrinsic deformation. However, to what extend they reflect the
dynamics of realistic nuclei remains a question. The oblate intrinsic
deformation observed in our models seems to be inconsistent with the
prolate dominance in real nuclei (see discussion in Sec. \ref{sec:Summary}),
moreover, the semi-magic nuclei with one type of valence nucleons
are generally not deformed. To attend to these issues we examine a
realistic valence space consisting of the $0f_{7/2}\text{ and }1p_{3/2}$
single-particle levels, allowing for both protons and neutrons. The
matrix element of the quadrupole operator for this model are constructed
using the harmonic oscillator single-particle wave functions, we use
the same effective charge for both types of nucleons. The multipole
operator in this form facilitates comparison with the SU(3) group. 

\begin{figure}
\includegraphics[width=3.4in]{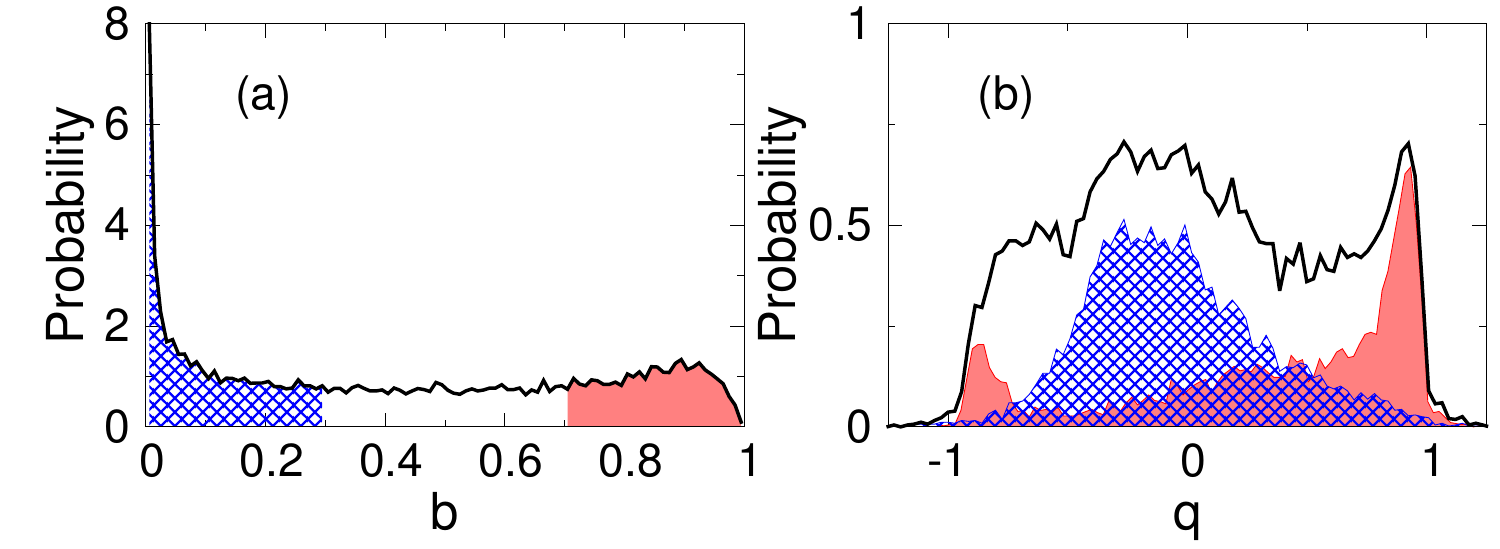}\caption{(Color online) $(0f_{7/2},1p_{3/2})^{8}$. (a) The distribution of
the fractional collectivity $b$. (b) The distribution of the intrinsic
quadrupole moment $q$. The solid black line outlines the probability
distribution for 31\% of realizations with the $0_{g.s.}$ state followed
by the $2_{\boldsymbol{1}}$ first excited state, both states with
isospin $T=0.$ The 8.8\% of realizations are collective and the 12.8\%
are non-collective. This figure is analogous to Figs. \ref{fig:jl19N6},
\ref{fig:multipole19N6}, \ref{fig:qqj19N8}, \ref{fig:ppvj13N6}-\ref{fig:ppvj13N22},
and the same shading is used in these figures. }
\label{fig:0f71p3}
\end{figure}

In Fig. \ref{fig:0f71p3} we present our results for the $(0f_{7/2},1p_{3/2})^{8}$
system with 8 nucleons: 4 protons and 4 neutrons. This corresponds
to the configuration space of $^{48}\text{Cr}$ nucleus. In Fig. \ref{fig:0f71p3}(a),
where the fractional collectivity $b$ is shown, a noticeable peak
that corresponds to collective realizations is observed. The distribution
of the quadrupole moment in Fig. \ref{fig:0f71p3}(b) shows prolate
and oblate peaks. The peaks are especially clear for the collective
realizations (shaded in uniform red). The non-collective cases in
Fig. \ref{fig:0f71p3}(b) are distributed around $q=0$ (shaded in
blue pattern). In agreement with the results in Ref. \cite{Horoi:2010},
in this TBRE the prolate intrinsic shape is more probable, as evident
from a bigger prolate peak. 

\begin{figure}
\includegraphics[width=3.4in]{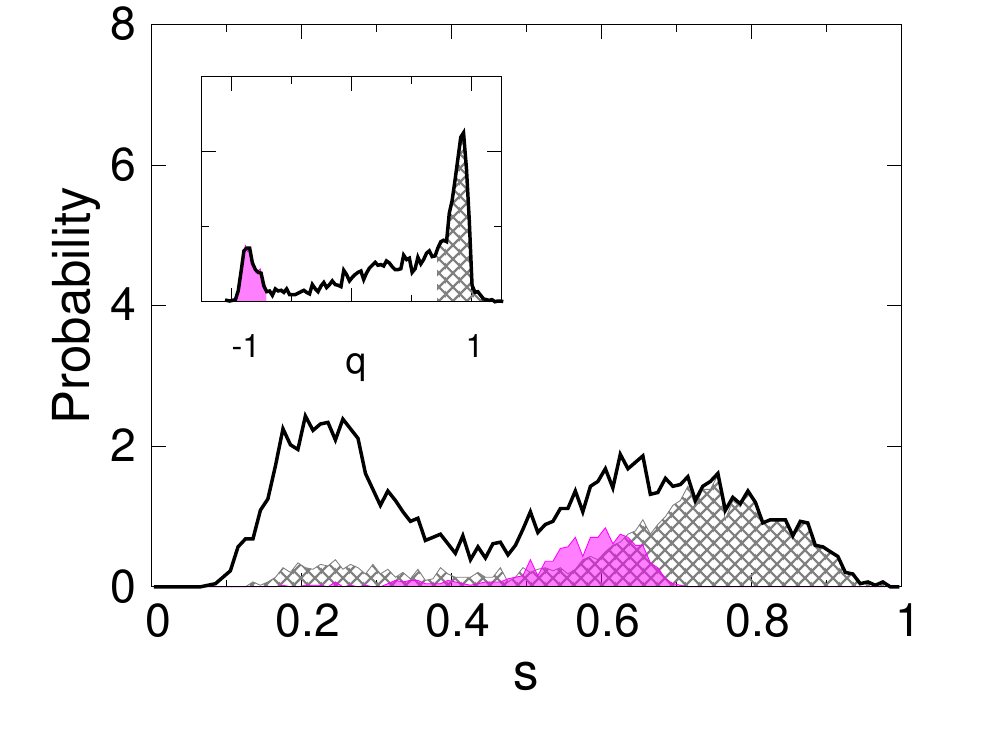}\caption{(Color online) $(0f_{7/2},1p_{3/2})^{8}$. The distribution of the
relative transition strength $s$ for the collective realizations
(shaded with uniform red in Fig. \ref{fig:0f71p3}). The 3.6\% of
prolate cases and 1.0\% of oblate are identified with shades of color
and pattern (see the inset).  This figure is analogous to Fig. \ref{fig:jl19N6_COL},
and the same shading is used as in Figs. \ref{fig:jl19N6_COL}-\ref{fig:jl19N6_trix}.}
\label{COLfig:0f71p3}
\end{figure}

In Fig. \ref{COLfig:0f71p3} we focus on the 8.8\% of realizations
that are collective. The quadrupole moments in Fig. \ref{fig:0f71p3}(b)
are further separated into prolate $q>0.7$ and oblate $q<-0.7$ cases
as shown in the inset of Fig. \ref{COLfig:0f71p3}. The same shading
is used in the main figure showing the distribution of the relative
transition strength $s$. The maximum possible value $s=1$ is reached
when the ground state wave function of the randomly selected Hamiltonian
coincides with that of the QQ Hamiltonian. From the summary in Tab.
\ref{tab:QQ} one finds that the $J=0,\, T=0$ ground state of the
QQ Hamiltonian, for which $s=1$, is prolate in this valence space.
Indeed, the distribution of prolate realizations is peaked at around
$s=0.8,$ while the oblate shapes have $s$ near $s=0.6.$

\begin{figure}
\includegraphics[width=3.4in]{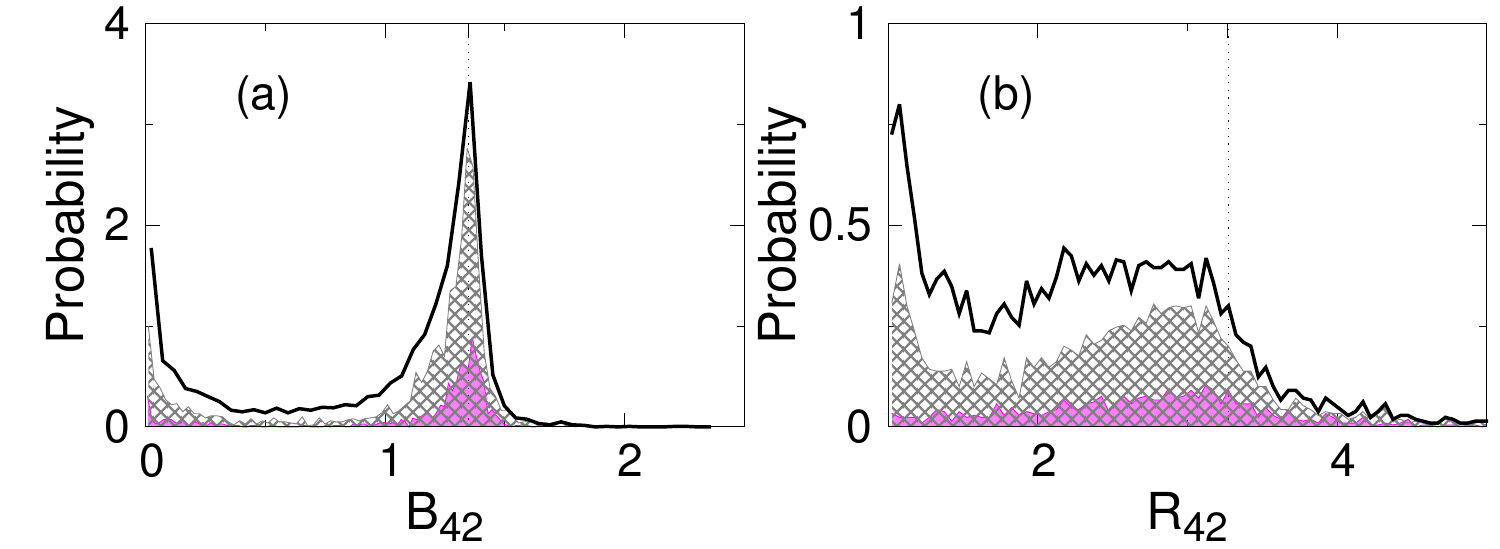}\caption{(Color online) $(0f_{7/2},1p_{3/2})^{8}$. (a) The distribution of
the deexcitation ratio $B_{42}$ defined in Eq. (\ref{eq:DX}). (b)
The distribution of the excitation energy ratio $R_{42}$ defined
in Eq. (\ref{eq:RR}). Collective realization discussed in Fig. \ref{fig:0f71p3}
are selected and, in addition, we require that the second excited
state has spin $4$. The fraction of such cases is 4.2\%, with 2.4\%
being prolate and 0.6\% being oblate, they are shaded separately with
the same patterns as in Fig. \ref{COLfig:0f71p3}. The values for
$B_{42}$ and $R_{42}$ from the QQ Hamiltonian listed in Tab.\ref{tab:QQ}
are shown with the vertical grid lines. This figure is analogous to
Fig. \ref{fig:jl19N6_COL024}, and the same shading is used as in
Figs. \ref{fig:jl19N6_COL}-\ref{fig:jl19N6_trix} and \ref{COLfig:0f71p3}.}
\label{fig:0f7p3_COL024}
\end{figure}

The distributions of $B_{42}$ and $R_{42}$ for collective realizations
are shown in Fig. \ref{fig:0f7p3_COL024}. This figure can be compared
to Fig. \ref{fig:jl19N6_COL024}. In both figures we use the same
shading to separate the prolate and oblate collective cases. In contrast
to Fig. \ref{fig:jl19N6_COL024}(a), both prolate and oblate realizations
in Fig. \ref{fig:0f7p3_COL024}(a) have a band structure with the
deexcitation ratio $B_{42}$ that is consistent with the rotational
value 10/7. This ensemble, based on the more realistic model space,
appears to have an energy spectrum that is closer to the rotational
spectrum. The distribution of $R_{42}$ in Fig. \ref{fig:0f7p3_COL024}(b)
is broad, but it has a peak around the rotor value of 10/3. 

\begin{figure}
\includegraphics[width=3.4in]{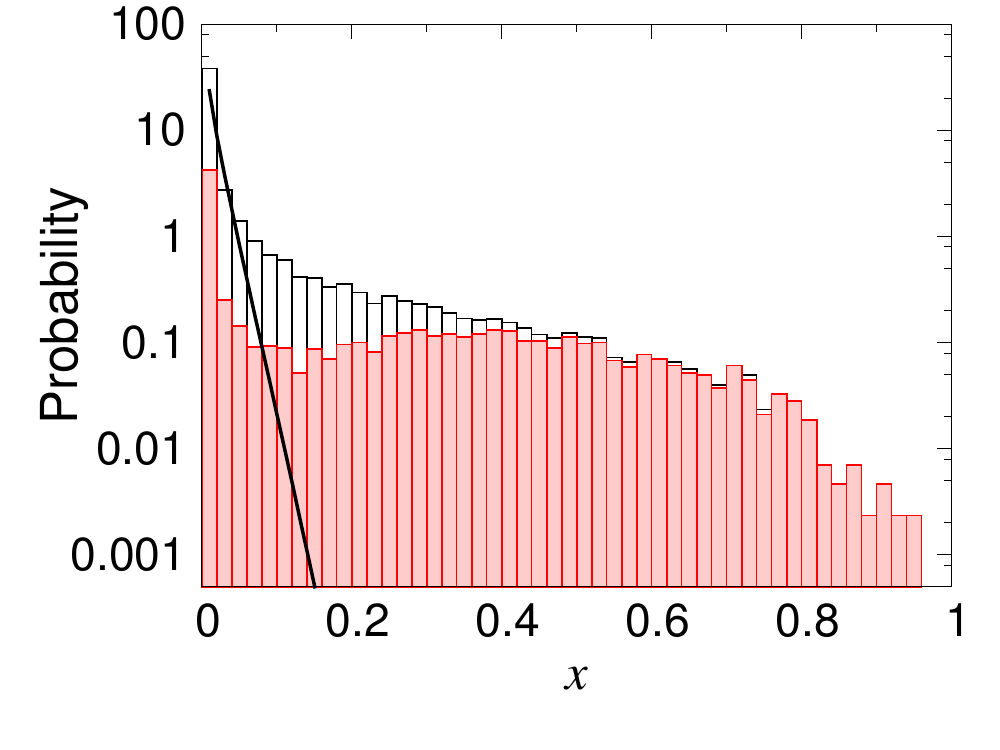}\caption{(Color online) $(0f_{7/2},1p_{3/2})^{8}$. The distribution of the
overlap $x$ defined in Eq. (\ref{eq:x}). The results for all $J_{gs}=0,T_{gs}=0$
states are unshaded; the fraction of such realizations is 56.3\%.
Collective realization that in addition to the $0_{gs}$ state have
the $T=0$, $2_{1}$ first excited state and $b>0.7$ are shaded (their
fraction is 8.8\%). Solid line shows the Porter-Thomas distribution,
which is expected for the overlap between uncorrelated states. }
\label{fig:overlap}
\end{figure}

As concluded in Ref. \cite{Horoi:2010}, realizations with rotational
features appear in random ensembles due to correlated interaction
matrix elements. Similarly to the single $j$ level model, it is natural
to attribute this collectivity to the QQ component in the Hamiltonian.
The overlap $x$ between the ground state wave functions of the two-body
random ensemble $|0_{gs}({\rm TBRE})\rangle$ and the fixed ground
state wave function of the QQ Hamiltonian is defined as\begin{equation}
x=|\langle0_{gs}(\text{TBRE})|0_{gs}(\text{QQ})\rangle|^{2}.\label{eq:x}\end{equation}
Fig. \ref{fig:overlap} shows the distribution of the overlap $x$
in the $(0f_{7/2},1p_{3/2})^{8}$ model. A similar approach has been
used in investigations of pairing coherence in random ensembles, see
review in Ref. \cite{Zelevinsky:2004}. We select 56.3\% of realizations
where the ground state quantum numbers are $J_{gs}=0$ and $T_{gs}=0$,
the ground state of the QQ Hamiltonian has the same spin and isospin.
The distribution of $x$ shown in Fig. \ref{fig:overlap} is compared
with the Porter-Thomas $\chi^{2}$ distribution. The latter emerges
for uncorrelated wave functions in the 126-dimensional space spanned
by the $J=0,T=0$ wave functions. As shown in Fig. \ref{fig:overlap}
the Porter-Thomas distribution drops abruptly, thus predicting that
cases with large $x$ are extremely unlikely. According to the Porter-Thomas
distribution the probability to find $x>0.1$ is only $0.03\%$ whereas
in the TBRE $x>0.1$ in $18.8\%$ of random realizations. To emphasize
the relation between the collective structure and the large QQ component
of the wave function we show in Fig. \ref{fig:overlap} the histogram
for collective realization (with states $0_{gs}$ and $2_{{\bf 1}}$
and $b>0.7$). It is clear that the collective transitions and rotational
structure emerge when the component of the wave functions that corresponds
to the eigenstate of the QQ Hamiltonian is large. 

\section{Summary\label{sec:Summary}}

Our studies show that a collective behavior that resembles realistic
is quite likely to be present in the ensemble with two-body random
interactions. This behavior appears to emerge due to the quadrupole-quadrupole
interaction component in the Hamiltonian. This component, as well
as some higher multipoles can establish some noticeable coherence
despite the overall many-body randomness and complexity. Similarly
to the moment-of-inertia-like $J^{2}$ term (that is responsible for
the ground state configurations with the maximum possible spin) the
QQ component, while not a constant of motion, is dynamically prevailing.
Let us list the supporting arguments:
\begin{itemize}
\item The fraction of random realizations that are quadrupole-collective
is extremely large as compared to the statistically expected number. 
\item In the two-body random ensemble, the quadrupole collectivity displayed
by the transition rates disappears when the QQ component in the interaction
is removed, see Fig. \ref{fig:multi_j19N8_JJ}. 
\item From investigations in Fig. \ref{fig:overlap}, as well as indirectly
form Figs. \ref{fig:jl19N6_COL}, \ref{fig:COLqqj19N8}, and \ref{COLfig:0f71p3},
it follows that the collective states in the TBRE have structure similar
to that of the QQ Hamiltonian eigenstates. 
\item In order to examine the shape and other quantitative characteristics
of the deformed mean-field we turn to the QQ Hamiltonian, for which
the geometry of the configuration space is the only parameter. The
values of the quadrupole moments, transition rates, and level spacings
for the models discussed in this paper are summarized in Tab. \ref{tab:QQ}.
In all cases the QQ Hamiltonian has a low-lying rotational spectrum.
The type of the quadrupole deformation and most of the other quantitative
measures in Tab. \ref{tab:QQ} are consistent with those observed
in the TBRE. This again suggests that the collective features seen
in the TBRE arise from the coherent QQ component. 
\end{itemize}
\begin{table}[t]
\begin{tabular}{lcccc}
\hline 
 & $b$ & $q$ & $B_{42}$ & $R_{42}$\tabularnewline
\hline
$(19/2)^{6}$ & 0.97 & -0.979 & 1.42 & 3.31\tabularnewline
$(19/2)^{8}$ & 0.95 & -0.969 & 1.43 & 3.27\tabularnewline
$(13/2^{+},13/2^{+})^{6}$ & 0.98 & -0.977 & 1.41 & 3.29\tabularnewline
$(13/2^{+},13/2^{-})^{6}$ & 0.98 & -0.977 & 1.41 & 3.29\tabularnewline
$(0f_{7/2},1p_{3/2})^{8}$  & 0.97 & 0.996 & 1.35 & 3.27\tabularnewline
\hline
\end{tabular}\caption{Characteristics of the QQ Hamiltonian. Listed in the table are the
values of the fractional collectivity $b$, quadrupole moment $q$,
ratios of the transition rates $B_{42}$ and the ratios of excitation
energies $R_{42}$. The models are the same as those considered in
our study of the TBRE. }
\label{tab:QQ}
\end{table}

Practically all deformed nuclei in nature are known to have a prolate
ground state shape. This prolate dominance has been widely discussed
in the literature \cite{Castel:1976,Castel:1990,Tajima:2002,Hamamoto:2009}.
An effort to pinpoint the origin of the phenomenon using the shell
model approach with random interactions is presented in Ref. \cite{Horoi:2010}.
While in this work we do not explicitly pursue the question of prolate
dominance, we are compelled to comment on the issue from the standpoint
of our findings. Our studies fully confirm the results in Ref. \cite{Horoi:2010}.
However, conclusions supporting the prolate dominance are difficult
to draw, instead we offer several observations. 

First, the quadrupole collectivity seen in the TBRE is due to the
QQ component in the Hamiltonian. This interaction and the geometry
of the valence space determine the deformation type. Thus, some questions
of the shape systematics can be addressed by considering the QQ Hamiltonian
and without invoking random interactions.

Second, the shape is determined by the valence configuration and by
the positions of the single particle levels. The role of the single-particle
level structure discussed by Hamamoto in Ref. \cite{Hamamoto:2009}
is possible to pinpoint using the TBRE as well as using analytic models,
e.g. the seniority model and Elliot's SU(3) model \cite{Abramkina:}. 

Third, due to particle-hole symmetry, which does not need to be exact,
the number of prolate and oblate configurations is approximately the
same within a given valence space. The deviations from this symmetry
affect only a few mid-shell systems where the two shapes compete.
The effect of the particle-hole symmetry is seen in our results in
Figs. \ref{fig:ppvj13N6} and \ref{fig:ppvj13N22}. 

To conclude, in this work we examined the quadrupole collectivity
that emerges in systems with two-body random interactions. A low-lying
spectrum, characteristic of a rigid rotor, is commonly observed. The
transition $B(E2,0_{gs}\rightarrow2_{1}),$ the quadrupole moment
of the first $2_{1}$ state, and the deexcitation ratio $B(E2,4_{1}\rightarrow2_{1})/B(E2,2_{1}\rightarrow0_{gs})$
are all consistent with that of the deformed rotor. A weak triaxiality
is also identified. The coherent dynamical role of the quadrupole-quadrupole
interaction component is established as a source of this behavior. 

The authors are thankful to V. Zelevinsky and J.M. Allmond for motivating
discussions. Support from the U. S. Department of Energy, grant DE-FG02-92ER40750
is acknowledged. The computing resources were provided by the Florida
State University shared High-Performance Computing facility.

\bibliographystyle{apsrev}

\end{document}